\documentstyle[12pt,aaspp,epsf]{article}

\begin{document}
\def\rpcomm#1{{\bf COMMENT by RP:  #1} \message{#1}}
\def\uptilde{\mathaccent"164}
\def\etal{{\it et al.~}}

\title{A Counter-rotating Bulge in the Sb Galaxy NGC~7331} 

\author{F. Prada$^{1}$}
\author{C.M. Guti\'errez$^{1,2}$}
\author{R.F. Peletier$^{3,1}$}
\and
\author{C. D. McKeith$^{4}$}

\affil{$^{1}$Instituto de Astrof\'\i sica de Canarias, 38200 La Laguna,
Tenerife, Spain}
 
\affil{$^{2}$University of Manchester, Nuffield Radio Astronomy Laboratories, 
Jodrell Bank, Macclesfield, Cheshire, SK11~9DL, UK}

\affil{$^{3}$Kapteyn Laboratorium, P.O. Box 800, 9700~AV~~Groningen, The 
Netherlands}

\affil{$^{4}$Department of Pure and Applied Physics, Queen's University of
Belfast, Belfast BT7~1NN, UK}

\journalid{Vol}{Journ. Date}
\articleid{start page}{end page}
\paperid{manuscript id}
\cpright{type}{year}
\ccc{code}
\lefthead{Prada et al.}
\righthead{A Counter-rotating Bulge}

\begin{abstract}

We have found that the bulge of the large, nearby Sb galaxy NGC~7331 rotates
retrograde to its disk. Analysis of spectra in the region of the near-IR
Ca II triplet along the major axis shows that, in the radial range between
5$''$ and $\sim$20$''$, the line of sight velocity 
distribution of the absorption lines  
is has two distinct peaks, and can be decomposed
into a fast-rotating component with v/$\sigma$ $>$ 3, and a slower rotating,
retrograde component with v/$\sigma$ $\sim$ 1 -- 1.5. 
The radial surface brightness
profile of the counter-rotating 
component follows that of the bulge, obtained from a
2-dimensional bulge-disk decomposition of a near-infrared $K$-band image,
while the fast rotating component follows the disk. At the radius
where the disk starts to dominate the isophotes 
change from being considerably boxy to very disky.

Although a number of spiral galaxies have been found that contain cold,
couterrotating disk, this is the first galaxy known to have a boxy, 
probably triaxial, fairly warm, counter-rotating component, which is 
dominating in the central regions. If it is a bar seen end-on, this bar
has to be thicker than the disk. We find that NGC~7331, even though it is 
a fairly early-type spiral, does not have a conventional, co-rotating bulge.
The fact that the inner component is retrograde makes us believe that it 
was formed from infalling material, in either stellar or gaseous form
(e.g. Balcells \& Quinn 1990). Another possibility however is that
the structure has been there since the formation of the galaxy. In this case
it will be a challenge to explain the large change in orientation of
the angular momentum when going outward radially.

\end{abstract}


\section{INTRODUCTION}

In the last decade many elliptical galaxies have been
found which display complicated multiple-component
structure in their kinematics  and, in some cases, also in their photometry.
This structure has been interpreted as being due to 
mergers (e.g. Balcells \& Quinn 1990), to components which were formed 
later from 
processed gas (Bender \& Surma 1992), or as triaxial systems in projection 
(Statler 1991). Sometimes these multiple components give rise to
disky isophotes, so that they can be found using photometry.
However in general they are much better seen in the analysis of the 
line-of-sight velocity distribution (LOSVDs). Examples of galaxies 
with multiple components can be found in e.g.
Franx \& Illingworth (1988), Rix \& White (1992) and Bender \etal (1994).
Typically they consist of a bright, pressure-supported body, and a fainter,
fast-rotating central disk, although this is not always the case 
(e.g. Balcells \& Carter 1993).

The kinematics of stars in spiral galaxies and lenticulars 
has been much less studied.
Bulges are generally thought to be supported by rotation (Kormendy 1993)
and no non-rotating bulge has been found up to now,
except maybe for the very small bulge of NGC~4550 (Franx 1993).
They rotate in the same direction as the disks, although
in general they are also partly pressure-supported. 
Recently however, a number of cases has been found in
which a fraction of the stars in the disk rotates in a retrograde way.
These include NGC 4550 (Rubin \etal 1992, Rix \etal 1992), 
NGC 7217 (Merrifield \& Kuijken 1994) and NGC 3593 (Bertola \etal 1996).
These counter-rotating disks are thought to have originated from
infalling gas, accreted in retrograde orbits in a plane close to
the equatorial plane (Merrifield \& Kuijken 1994).

In this paper we present the discovery of a different type of galaxy: 
one in which the bulge
rotates retrograde to the disk. We have found this by analyzing LOSVDs
along the major axis of the large, nearby 
Sb (Sandage \& Tammann 1981) galaxy NGC~7331. Here two components are seen,
of which the slow, counter-rotating one corresponds to the boxy bulge 
of this galaxy. 
Explaining the formation of a system like this will
be a challenge to galaxy formation theories. Section 2
of this paper gives details about the observations. In  Sections 3 and 
4 we describe our analysis of morphology and  the LOSVDs. The discussion 
 and conclusions are given in  Section 5.

\section{OBSERVATIONS AND DATA REDUCTION} 

We obtained long-slit spectra 
with the ISIS spectrograph on the 4.2m WHT at La Palma (August 1992)
along the major axis (p.a. 167$^{\circ}$, Bosma 1981) of NGC 7331. The 
slit was centered
on the apparent optical nucleus, with a projected slit width of $1''$, 
slightly undersampling the $1''$ seeing. The spectral dispersion was 
2.1 pixels of 25 km~s$^{-1}$~pixel$^{-1}$ each. The spectrum, of 1800s, 
was taken in
the red arm of ISIS, with the wavelength 
centered at 8700 \AA\, to include  the CaII IR triplet (8494, 8542, 8662 \AA).
The bias subtraction, flat-fielding  and 
wavelength calibration of the data were carried out using  FIGARO.
Sky lines were subtracted using template 
spectra from the edges  of the slit. As a stellar template star we 
observed the K0 giant HR 5631. To confirm
and strengthen our results we have also analyzed a  spectrum
from the La Palma Archive,  of NGC~7331 along p.a. 172$^{\circ}$. These data 
were taken with the same instrumental setting and exposure time  
in September, 1991, by 
D. Carter and M. Balcells. Their seeing FWHM was 1$''$ and their  spectral 
resolution was  2.5 pixels FWHM. From the same observing run we used 
the G8 giant HR 7753 as a  template.

To study the morphology, we took images in the Cousins $I$-band and the
infrared $K$-band. The $I$-band image was taken in June 1990 at the INT at
La Palma  with the 2.5~m  INT and a 590 $\times$ 400 EEV CCD. 
The pixel size was 0.549$''$, with a seeing of $\approx$ 1.2$''$.
The $K$-band data were obtained at UKIRT in June 1994 using IRCAM3
with a 256 $\times$ 256 InSb detector. A mosaic was made  to have 
a total field of 80$''$ $\times$ 80$''$, with pixels of 0.291$''$. The 
data were taken under photometric conditions with a seeing 
of $\approx$ 0.9$''$. Details of the data reduction can be 
found for the optical data in Balcells \& Peletier (1994) and for the 
infrared data in Peletier \& Balcells (in preparation).

\section{THE STRUCTURE OF NGC~7331}

To determine the structure of the central regions of NGC~7331 we fitted
ellipses to the images in the $K$- and $I$-band using Galphot
(see J\o rgensen \etal 1992). Fig.~1 shows the surface brightness profiles
in $I$ and $K$ as a function of major axis radius.
There are some minor differences in the inner 60$''$
between both profiles, mostly due to extinction by dust.  
Because of the smaller size of the $K$-band image 
it was not possible to extend the fit as far as in the $I$-band.
Consequently the slope of the profile in $K$ in its outer parts
is much steeper than the slope of the profile in $I$ at large $I$-band radii.  
If one tries to decompose the galaxy into an exponential disk and a 
r$^{{1\over{4}}}$ bulge, only based on the surface brightness profile,
one will get a much larger bulge in $I$ than in $K$ (see Fig.~1). 
In galaxies that are not face-on however, one can also use the information
available in the axis ratio distribution, and in this way try to isolate
a flat disk and a rounder bulge (Kent 1985). 
Applying Kent's decomposition method for this galaxy
of inclination 75$^{\rm o}$
we have obtain in both bands the fit that was obtained in $K$ using the 
previous method. The reason for this is that the ellipticity in this galaxy 
only changes up to $r$ = $\sim$ 10$''$, so that the only possible
solution can be a small bulge and a large inclined disk.
In Fig.~2 we have plotted
ellipticity, major axis position angle and boxyness C4 as a function of
major axis radius. 

It seems that the morphology shows three large components: the central
bulge, with effective radius of $\sim$ 10$''$ ($K$), and  then 
two flat components: an intermediate 
component with scale length of $\sim$ 25$''$,
and from 60$''$ onwards the large  outer disk. 

Looking in more detail at the position angle and C4 major axis profiles  
we see in the inner 5$''$ a boxy 
structure (C4=--0.017 at 2.5$''$) and a position angle twist 
of 10$^{\rm o}$  between
2.5$''$ and 10$''$. This component is entirely part of the bulge.
Between 5$''$ up to 15$''$ the morphology is suddenly very disky, with
C4 values $\sim$+0.04 which are  much larger than the normal values for disky 
elliptical galaxies (e.g. Bender \etal 1988). 
Beyond  15$''$, C4 depends on passband, and is probably severely affected
by extinction. At larger radii it stabilizes to a value of $\sim$+0.02.
No major changes of position angle and ellipticity are seen beyond 10$''$.

\section{THE LINE-OF-SIGHT VELOCITY DISTRIBUTION IN NGC~7331}

To obtain the LOSVDs, we compared  
the shift and Doppler broadening of the  CaII IR triplet region with
a standard reference stellar spectrum (K0III) recorded the same night and with
the same instrumental configuration as for the galaxy. We used an extension
of the unresolved Gaussian decomposition technique  developed by 
Kuijken \& Merrifield (1993). The basic difference between their and our
algorithm is that we perform a two-dimensional fit to the long-slit
spectra by using  unresolved Gaussian 
components with dispersion 2 pixels and separation  3 pixels in both spatial
and spectral direction. 
In general the results of both algorithms
are the same, but ours should be more robust to
spurious features in regions with low signal to noise ratios.
 A detailed description of the algorithm  will be given 
in a separate paper (Guti\'errez
\& Prada, in preparation). 
In Plate~1, a surface plot of the LOSVDs along the 
major axis of NGC~7331 is shown. 
The LOSVDs in the inner 4$''$ are 
nearly symmetric (see Fig. 3, top panel). Between 4$''$ and 7$''$ on 
both sides of the nucleus a strong asymmetry is evident towards the 
systemic velocity (see Fig. 3, middle panel)
of the galaxy. Between 7$''$ and 15$''$ the LOSVDs show  two 
separate peaks, with the fainter one  crossing  the systemic 
velocity (see Fig. 3, bottom panel).  Further out, the 
fainter  peak disappears and the LOSVDs are again defined by a single  
component. The archive 
spectrum   was analyzed in the same way, and the same features 
were found.  
We have fitted one, and where possible two Gaussians to the LOSVDs, and 
display the kinematic parameters
(recession velocity, velocity dispersion and flux) as a function of 
major axis radius in Fig.~4. Because of the way our algorithm works
the individual points are slightly correlated; the algorithm effectively
smoothes the data in this figure 
in the spatial direction with a Gaussian of FWHM 1.6''.
We find that between 5$''$ and 15$''$ on 
both sides of the nucleus a fraction of the 
stars rotates slowly and retrograde to the rest of the 
galaxy. In the inset of Fig.~1 we have plotted the relative intensity
of both components, obtained from the LOSVDs. The profile
of the fainter component agrees well with that of the bulge, as obtained
from the photometric decomposition in K. In the inner 5$''$, since an 
unambiguous decomposition of the line profiles was not possible due to lack 
of resolution, we have fixed the velocity of the bulge component to the
velocity of the dashed line in Fig.~4, as well as the intensity ratio 
of both components, which we assumed to be the ratio of the components 
from the photometric
decomposition. It was only possible to obtain reasonable decompositions
if the dispersion of the low-velocity component was taken to be 
110 km/s or larger. For that reason its dispersion was fixed at 110 km/s.
The velocities obtained for the disk are plotted in the upper panel of Fig.~4
as filled circles.

\section{DISCUSSION AND CONCLUSIONS}

To summarize the observations: the inner parts of the galaxy consist
of a boxy component, dominating the inner 5$''$. It shows position angle 
twisting, rotates retrograde to the rest of the galaxy, and is rounder.
Outside 5$''$ the disk dominates. This component is much colder,
is probably flat, and responsible for the disky isophotes.
At appr. r=100$''$ the surface brightness profile becomes shallower.

The boxyness and position angle twist make it very likely that
we are seeing a triaxial object counter-rotating to the main body
of the galaxy. It could be that we have a configuration like in
NGC~4736 (M\"ollenhoff \etal 1995), where the bar is oriented
end-on towards us, or alternatively, that the galaxy has no bar, and that
we are seeing a triaxial bulge. The steep rise in the surface brightness
profile and the low ellipticity show that it can't be a classical, flat
bar with a Freeman-type profile (Freeman 1966). From this data we cannot
distinguish between these two alternatives any further. Whatever the 
case, the galaxy does not contain a large, co-rotating bulge; if it
has a bar a small bulge, as in NGC~4736, can be hidden.

It is likely that the origin of the central component is external.
Evidence for this would be the fact that it is retrograde, and maybe 
the boxyness. Simulations for elliptical galaxies have shown that it is 
possible to create a central, counter-rotating body as a result of a 
stellar merger (Balcells \& Quinn 1990). Another possibility is that
the central component is formed as a result of an instability
in the disk of counter-rotating stars, like the one of NGC~7217
(Merrifield \& Kuijken 1994). In that case however the counter-rotating
disk must have accreted from outside first. A third possibility might be
that the counter-rotating stars were formed at the formation epoch 
together with the rest of the galaxy. Up to now no models have been
proposed where two counter-rotating components were formed from
one protogalactic cloud, and where the angular momentum changes so
drastically as a function of radius. We might observe here something
similar to a counter-rotating bar within a bar, for which Friedli (1996)
has shown that it can be very longlived.

Further evidence for a triaxial potential in the center of this galaxy
comes from the H$\alpha$ velocity field by Marcelin \etal (1994). They
find the typical Z-shape in the isovelocity contours, characteristic
of non-circular potentials. The galaxy shows some low-level
activity (Ho \etal 1995). On larger scales, the motion of the 
gas is quite regular (von Linden \etal 1996, Begeman 1987).

We now ask ourselves how unique this galaxy is. 
Studies using e.g. the Fourier quotient
method (Tonry \& Davis 1979) generally find that the bulge is 
dynamically supported by rotation, and rotates in the same direction
as the disk (Kormendy 1993). If however the bulge to disk ratio
diminishes, it becomes difficult to find any slow-rotating component
in this way. To show this, we give the h3 and h4 components, as
defined by van der Marel \& Franx (1993), for
NGC~7331 in Fig.~3. They are similar to h3 and h4 profiles of  many
elliptical galaxies (e.g. Bender \etal 1994). Since however a somewhat
fainter, co-rotating component could give the same h3 and h4 as we
are seeing here, we need to investigate the line profiles in more detail.
By now, there are several spiral galaxies and 
S0's for which the line profiles have been studied, There are many 
cases with a hot bulge and a cold disk rotating in the same direction
(e.g. NGC 3115 (van der Marel \etal 1994), NGC 4736 (M\" ollenhoff \etal 1995),
NGC 4594 (Wagner \etal 1989)). But as far as we know, no galaxy has been 
found where the central and outer component rotate opposite to each other.
We conclude that NGC~7331 is special in this respect.

We would like to end advancing some speculations. First, we find here 
a very boxy, slowly rotating inner component, and a large, fast
rotating disk. Many giant elliptical galaxies on the other hand
have a boxy, even slower rotating, main body, and a small, rotating disk.
Apart from the relative scales, the similarity of both systems is
striking, and has to imply that the formation scenario of spirals
like NGC~7331 and ellipticals cannot be too different.
And secondly, we see a discontinuity in the slope of the surface brightness
profile at r=100$''$. Since this galaxy is of type Sb, it is rather
peculiar that it doesn't have a larger, co-rotating bulge. Could it
be that the component that dominates the light between 5$''$ and 100$''$
was the previous bulge? It however is flat, rotates fast, and is disky,
but according to Kormendy (1993) many late type spirals have 'bulges'
that look like this. This question obviously is still open, and needs
much more investigation.

\acknowledgements This research is based on observations obtained
at the WHT and INT, operated by the RGO at the Observatorio del
Roque de los Muchachos at La Palma of the IAC. We thank E. Simonneau,
E. Perez, M. Balcells, K. Kuijken, 
D. Fisher, M. Franx, R. Bottema and R. Sancisi
for helpful discussions. The spectroscopic  observations were performed on 
a service night (August 1992) by Dr. R. Clegg.  Part of the observations 
are based on the La Palma Archive of the Royal Greenwich Observatory. We 
thank the referee for very valuable suggestions, and T. Mahoney for his help 
with the text.

\setcounter{figure}{0}
\begin{figure}
\plotone{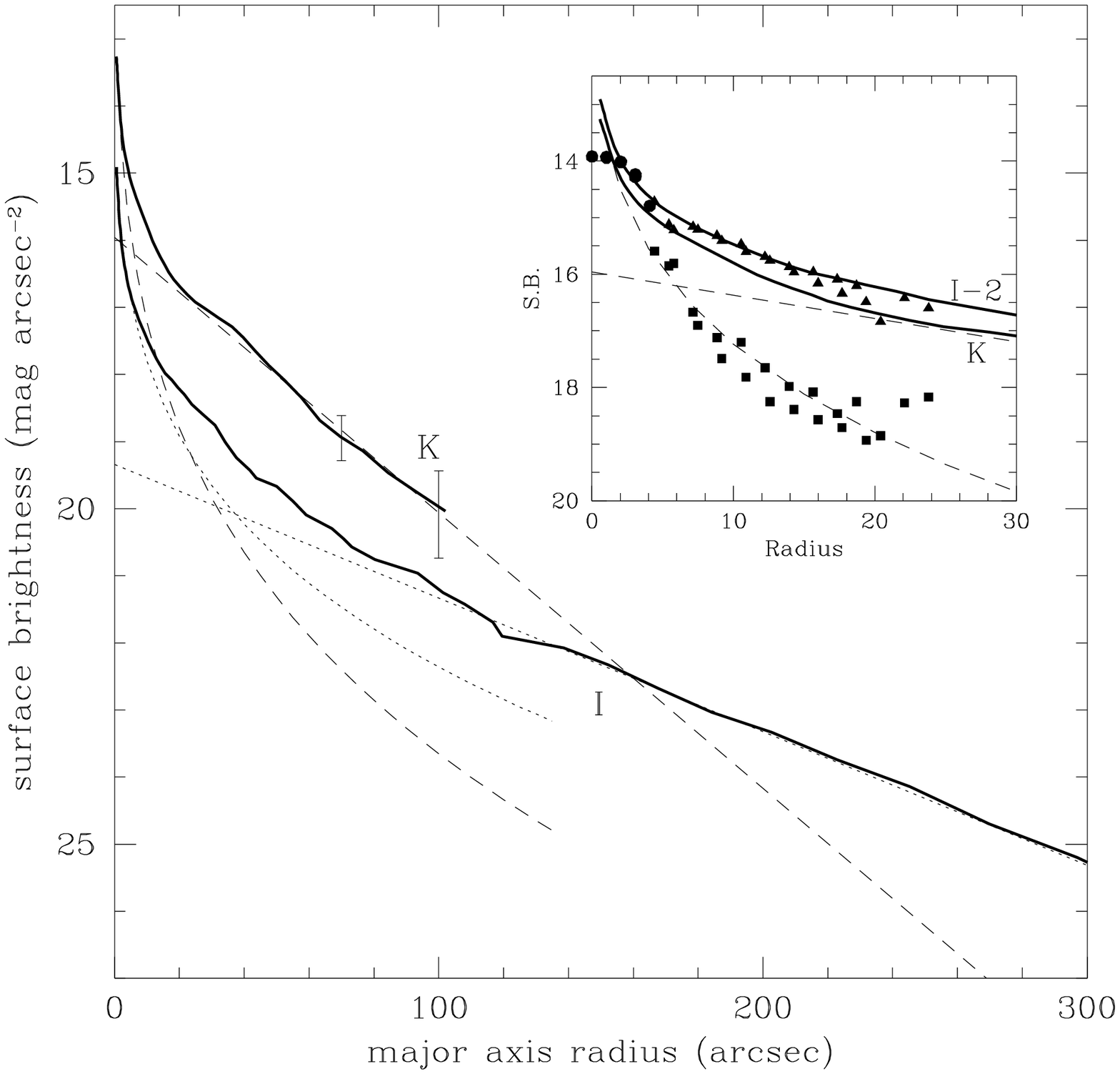}
\caption{
Plot of the major axis surface brightness profile of NGC~7331 in  
$I$ and $K$ (solid lines). Both profiles has been decomposed into an 
exponential disk 
and  an spheroidal r$^{1/4}$ component (dashed  and dotted lines for the
$K$ and $I$-band respectively). The  fit parameters  
are also shown. Also plotted are the  one-sigma error-bars 
due to uncertainties in the determination of the sky background. 
(1$''$$\sim$72 pc for a distance of 14.9 Mpc, see Begeman 1987).
In the inset we have enlarged the central regions. Here also plotted are
the fluxes of the high (triangles)  and low-velocity (squares) 
components, from the
decomposition of the line profiles, shifted by an arbitrary constant. The
I-band profile (minus 2 magnitudes) is also shown here. }
\end{figure}
\begin{figure}
\plotone{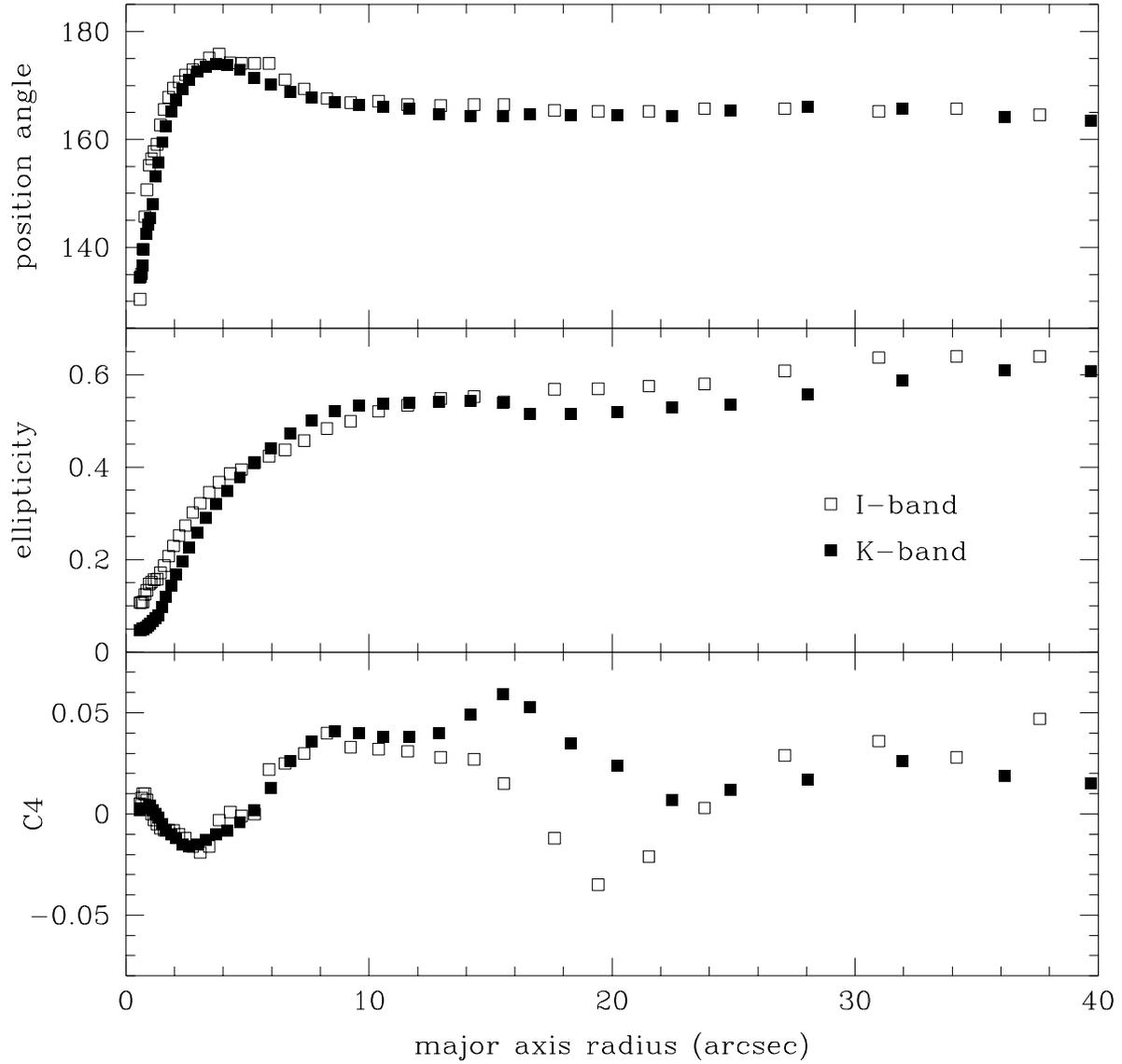}
\caption{
Isophotal analysis of the $I$- and  $K$-band surface 
photometry of 
NGC 7331. Top panel: the major axis position angle profile; 
 middle panel: the ellipticity profile; and the 
bottom panel: the profile of the C4 Fourier coefficient indicating deviations
from elliptical shape.}
\end{figure}
\begin{figure}
\plotone{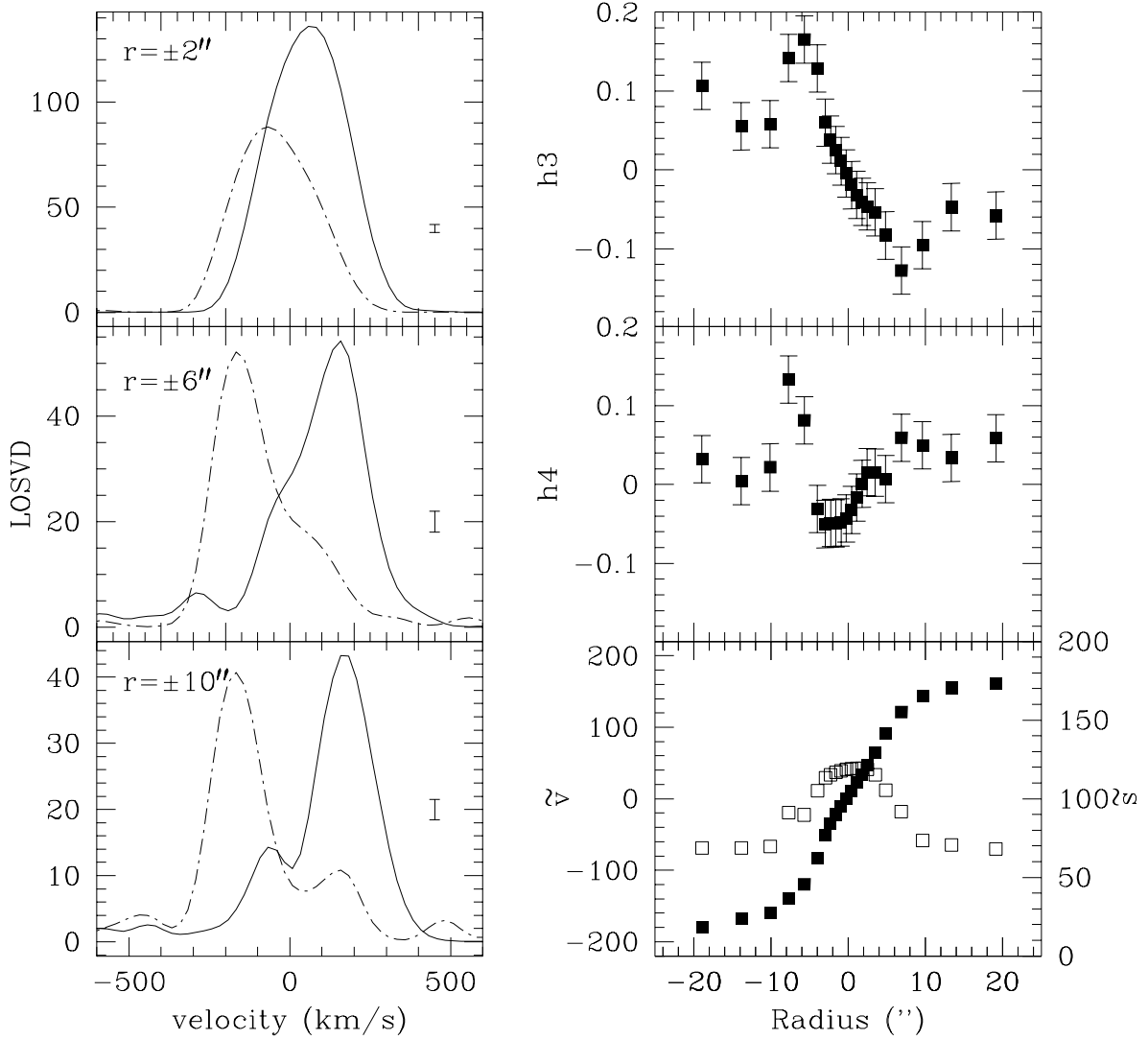}
\caption{On the left three LOSVDs 
along the major axis of NGC~7331 at different  distances from the center.
Dot-dashed line represent the NW side (blueshift) and full line the SE 
side (redshift). Typical error-bars are shown. On the right we give the
profiles of h3, h4, \~v (filled symbols) and  $\tilde \sigma$ according 
to the definition of
van der Marel \& Franx (1993).}
\end{figure}
\begin{figure}
\plotone{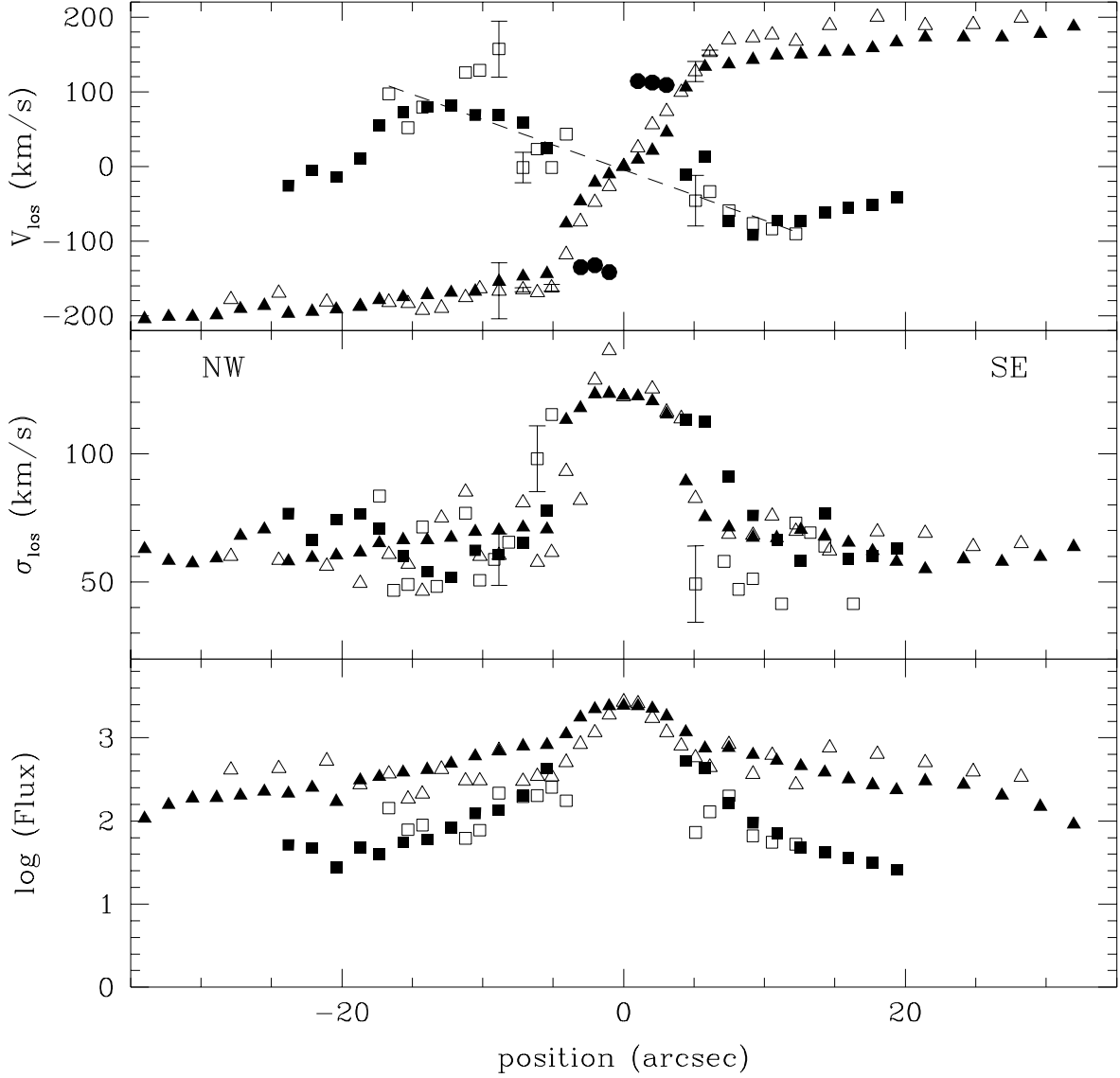}
\caption{ Kinematic parameters for  fitting two Gaussians to the LOSVDs 
along the major axis of NGC~7331. Open and filled symbols correspond to 
the independent spectra at p.a. 167$^{\circ}$ and p.a. 172$^{\circ}$ 
 respectively (our observations and those from the archive). Triangles and 
square symbols refer to the 
brighter and fainter components present in the 
LOSVDs. The errors in the fit to the LOSVDs are comparable in general to the 
size of the
symbols, larger errors are indicated. 
The dashed line in the top panel is 
a model for the bulge component. In the inner parts, where no unambiguous
decomposition was possible, we have fixed the velocity and dispersion 
of this component, as well as the flux ratio (see text). The high-velocity
component that is obtained in this way is indicated with filled circles.}
\end{figure}
\addtocounter{figure}{-4}
\begin{figure}
\epsfysize=13cm \epsfbox[045 182 367 611]{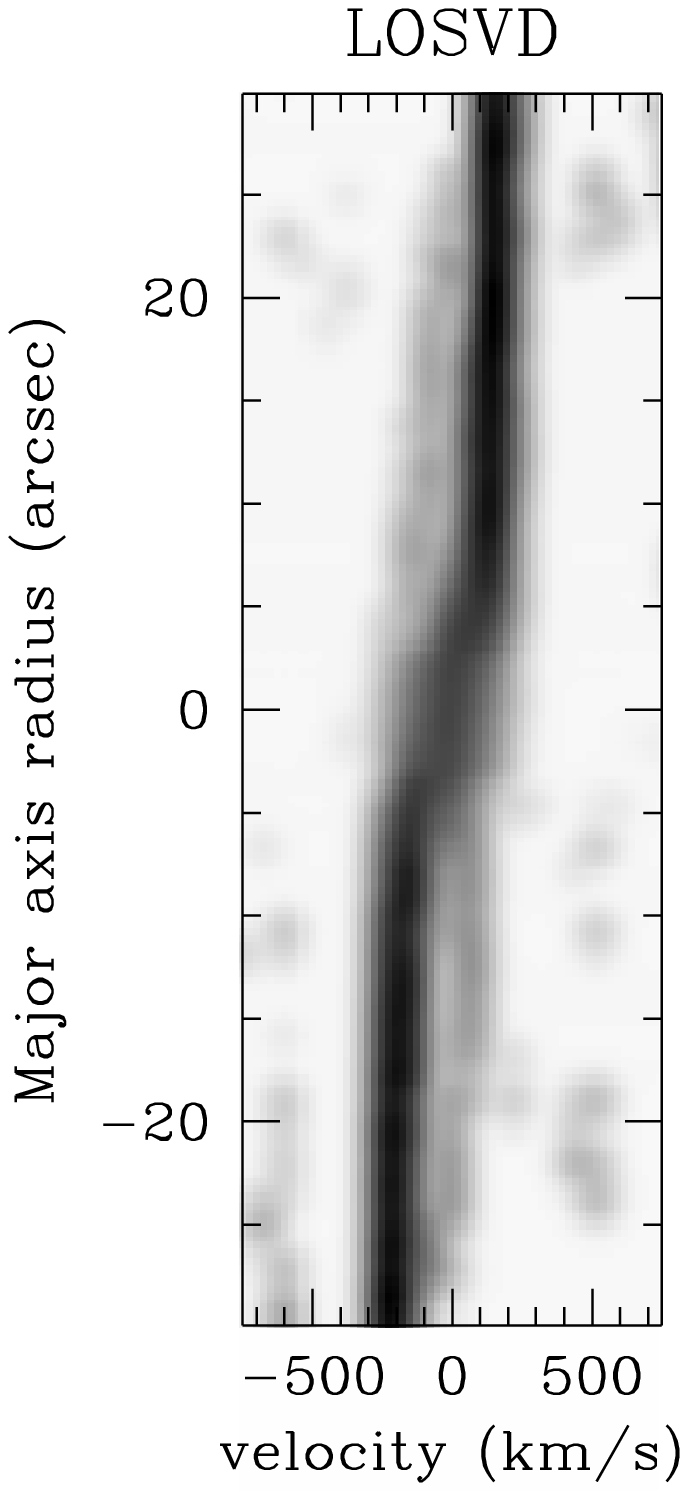}
\caption{
Plate 1: Greyscale plot of the normalized stellar line-of-sight velocity distribution
along the major axis of NGC~7331, where for presentation purposes the data
in the spatial direction has been smoothed with a Gaussian with a FWHM of 4$''$.}
\end{figure}

\end{document}